\documentclass{rtav}

\usepackage[colorlinks=true,pagebackref]{hyperref}
\usepackage[margin=1.25in,top=1in,left=1.25in,headheight=\baselineskip]{geometry}%

\graphicspath{{./figs/}}%

\begin{document}

\midspace

\begin{center}

\vspace*{5mm}

{\Large\bf\ctitle On the prediction of shrinkage defects by thermal criterion functions}%

\vspace*{5mm}

\end{center}%

\begin{center}

{\large R. Tavakoli\footnote{Department of Material Science and
Engineering, Sharif University of Technology, Tehran, Iran, P.O. Box
11365-9466,  Email:
\href{mailto:tav@mehr.sharif.ir}{tav@mehr.sharif.ir},
\href{mailto:rohtav@gmail.com}{rohtav@gmail.com}}%
}%

\end{center}%

\begin{center}%
{\small\it Last modification: November 19, 2009}%
\end{center}%


{\noindent\bf\csection Abstract. }%
The indirect prediction of shrinkage induced solidification defects
is considered in this study. The previously suggested criterion
function methods, in particular the Pellini and Niyama criteria are
analyzed in details, and their shortcomings are shown as a result of
our analysis (e.g. the scale/shape-dependency of critical values and
the inability to distinguish between cold- and hot-spots). To
moderate limitations related to criterion function methods, a new
method is introduced to predict the location of centerline shrinkage
in metal castings. Unlike the alternative methods which are derived
more empirically based on the result of experimental observations,
the suggested method in this study is derived theoretically based on
a heuristic two-scale, macro-meso-scale, approach. The application
of the suggested method is limited to low freezing range alloys. The
feasibility of the presented method is studied by comparing
numerical results against the available experimental data.%

{\noindent\bf\csection Keywords. }%
criterion function, defect prediction, macroshrinkage prediction,
multiscale analysis, Niyama criterion, two-scale approach.%

{\noindent\bf\csection PACS. }%
81.30.Fb, 91.60.Ed, 91.60.Hg

\newpage



\section{Introduction. }%
\label{sec:introduc}%

Porosity is one of the major defects in castings which results in a
deterioration of mechanical properties, in particular fatigue and
a reduction in ultimate tensile strengths. It is induced by two mechanisms,
solidification shrinkage and gas segregation, both of which occur
concomitantly but with different intensities \citep{campbell2003c}.

The goal of this study is two folds. First to analyze available
thermal criterion function methods for the prediction of shrinkage
defects in castings. Second, to introduce a new thermal criterion
function which annihilate some limitations related to alternative
criterion function methods. In practice, the formation of shrinkage
defects has a close relation to gaseous defects. Consequently, as
they are difficult to distinguish, in particular from the viewpoint
of defect nucleation, we shall attend to the mechanism of gas
porosity formation to explore validity conditions for our criterion.


\section{Spatial length scales involving the casting process. }
\label{sec:spscales}%

\begin{figure}[ht]%
\begin{center}%
\ifshowfigs%
\includegraphics[width=8.cm]{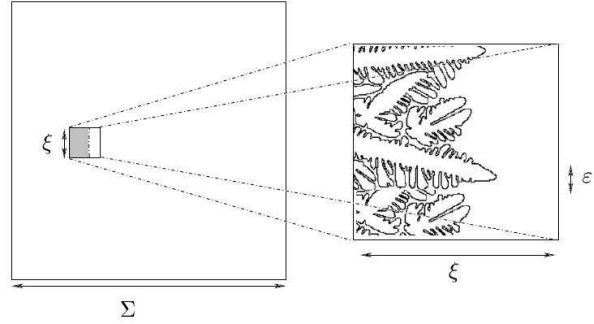}%
\fi%
\caption{Sketch of macro- ($\Sigma$), meso- ($\xi$), and microscale
($\varepsilon$) in a casting process. The images
also show the topology of the solid/liquid interface from a macroscopic (left) and mesoscopic (right) viewpoint.}%
\label{fig:castingscales}%
\end{center}%
\end{figure}

Alloy solidification is a multiscale phenomenon in nature. It
includes multiple spatial scales such as macro, meso, micro, atomic
and sub-atomic (electronic) scales. In the case of metal casting, it
is possible to consider a three-scale situation with respect to
space (see the sketch in \autoref{fig:castingscales}); i.e., there
exist

\begin{itemize}%

\item[$\bullet$] {\it macroscale}: a length $\Sigma\approx 1$--$10^{-2}$ m, in which the whole process takes
place;%

\item[$\bullet$] {\it microscale}: length scale $\varepsilon\approx 10^{-4}$--$10^{-6}$ m related to typical dendrite tip radii;%

\item[$\bullet$] {\it mesoscale}: a length $\xi$, $\varepsilon \ll \xi \ll \Sigma$, which
marks the finest resolution for the description that is of practical
importance in porosity formation. Since porosities mainly form as a
result of crystal interactions, the typical primary dendritic arm
spacing or grain size is a feasible choice for this length scale, i.e. $\xi\approx 10^{-2}$--$10^{-4}$ m;%

\end{itemize}%

\begin{figure}[h]%
\begin{center}%
\ifshowfigs%
\includegraphics[width=9.5cm]{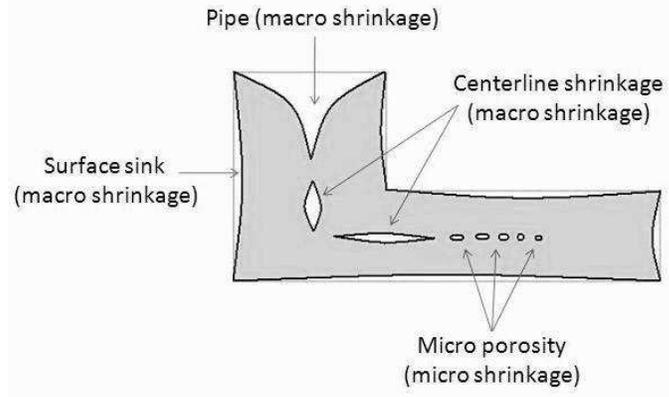}%
\fi%
\caption{Definition and classification of shrinkage defects within a casting ingot.}%
\label{fig:defect_types}%
\end{center}%
\end{figure}

It is common to classify shrinkage defects in the casting process
based on their physical appearance and spatial scales. Porosity in
castings can be classified by the size of the pores as
macroporosity and microporosity or dispersed porosity (see
\autoref{fig:defect_types}). Macroporosity defects can be divided
into three categories: pipe shrinkage, centerline shrinkage and
surface sink (also called external shrinkage porosity).

The spatial scales of macroporosity and microporosity are formally
matched with the above mentioned macro and meso scales,
respectively. Due to the limitation of available computational
resource, it is almost impossible to simulate a practical casting
process at the mesoscale. Consequently, we have to perform a
macroscale solidification analysis to predict solidification induced
defects. Therefore, only a fraction of defects can actually be
resolved by such a macroscale simulation. This spatial filter
usually covers the macro porosities and hopefully the upper-range of
micro porosities.

From a mesoscale viewpoint, in a typical casting process the
solid/liquid interface is non-planar as a result of unstable
solidification due to the melt supercooling in the vicinity of the
solidifying front and/or the constitutional supercooling in the case
of alloy solidification. However, from a macroscopic point of view,
it is possible to consider the solid/liquid interface as a sharp and
planar interface when we have a narrow mushy region. In this case
the mushy zone usually includes almost parallel short columnar. Some
authors (e.g. \cite{niyama1982msp,carlson2009prediction}) assumed
the flow of melt is governed by a very simple form of Darcy-law and
then used it to predict the susceptibility of cease of flow and so
formation of meso-scale solidification defects. However, we believe
that this is a blind assumption which ignores couple of important
physics behind such a meso-scale fluid flow and defect formation.
Later in this paper we illustrate our criticisms in this regard.

In the case of wide freezing range alloys, in particular when the
thermal diffusitivity of melt is large and the thermal conductivity
of the mold is small, the entire casting would quickly fill with a
mixture of solid dendrites and liquid, the remaining volume being
filled by the mushy material until the end of solidification. In
this solidification mode, it is not possible to distinguish between
solid and liquid phases from a macroscopic viewpoint. In this case
it is reasonable to consider the dendritic skeleton as a randomized
porous media like \cite{flemings1974sp}. Later, we shall argue that
using thermal criterion functions are not feasible for such cases.

\begin{figure}[ht]%
\begin{center}%
\ifshowfigs%
\includegraphics[width=8.cm]{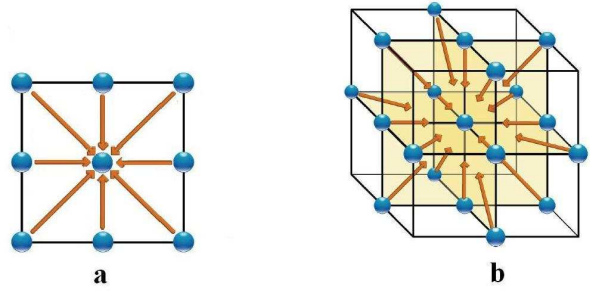}%
\fi%
\caption{Search directions to compute the Niyama gradient in a) 2D and b) 3D cases.}%
\label{fig:niyamadirections}%
\end{center}%
\end{figure}

Prior to closing this section, let us to define some quantities
which are used throughout this paper. The total freezing time of
casting, local freezing time, local cooling rate at the freezing
time, local velocity of the solidus isotherm and feeding gradient
are denoted by $t_s$, $t_f$, $R$, $v_s$ and $G$, respectively. The
feeding gradient referred to here is the component of the
temperature gradient in the feeding direction, i.e.
\begin{equation}%
\label{niyamagrad}%
    G = \nabla \theta_c \cdot \hat{\textbf{n}}%
\end{equation}%
where $\theta_c$ denotes the temperature field inside the cast
region and $\hat{\textbf{n}}$ is the unit normal of the solid/liquid
interface directed toward the liquid region. Since the iso-contours
of the local freezing time field represent the position of the
solid/liquid interface at a specific time, $\hat{\textbf{n}}$ satisfies
the relation $\hat{\textbf{n}}=\nabla t_f /
|\nabla t_f|$. An upwind-like method was suggested by
\citet{niyama1981psi} to compute the feeding gradient on a
structured grid; specifically, to compute $G$ on a cell $c$, we calculate%
\begin{equation}%
\label{niyamagrado}%
    G = \max_{\ i \in nb\ } \frac{\theta_i - \theta_c}{\ \|\textbf{x}_i - \textbf{x}_c\|_2}.%
\end{equation}%
where ${nb}$ denotes the
liquid neighborhood of the desired cell. In the case of two
dimensions, this neighborhood is limited to those cells among the 8 nearest
neighbors in which the temperature is larger than the solidus
temperature (see \autoref{fig:niyamadirections}). In  three spatial dimensions, this neighborhood
has 26 members.
Knowing the local freezing time field as a result of the solidification analysis,
$v_s$ satisfies $
   v_s = (\nabla t_f \cdot \hat{\textbf{n}})^{-1}%
$.
It can be computed in the same way as described above for $G$.

Since the feeding capacity of the liquid metal ceases when the
volume fraction of the solid phase is larger than a critical value,
some authors suggest to evaluate the solidification parameters at
a temperature $\theta_s^+$ slightly higher than the solidus temperature.
For example \citet{carlson2002dnf} suggested to take
$\theta_s^+$ equal to the solidus temperature plus 10 percent of the
interval between the solidus and liquidus temperatures, i.e.,
$\theta_s^+ = \theta_s+ 0.1 (\theta_l-\theta_s)$, where $\theta_l$
and $\theta_s$ denote the liquidus and solidus temperatures,
respectively. In the present study every parameter which is computed
at $\theta_s^+$ is denoted by a plus superscript, $\Box^+$.


\section{Related work. }%
\label{sec:relatedworks}%

During the past two decades several computational methods have been
developed to predict shrinkage defects in castings
\citep{lee2001mma,stefanescu2005css}. Early approaches
include the numerical solution of the energy equation and evaluation
of a local criterion function to predict shrinkage defect
susceptibility. The applied criterion was usually a function of $G$,
$R$ and $v_s$. These methods are known as criterion function
methods. More advanced approaches, e.g.
\citep{pequet2002mmm,wang2007stm,drenchev2007cmo}, include the
solution of energy and momentum equations coupled with a suitable
mesoscale pore nucleation and growth model. Although these methods
include more physics and are more accurate, none of them could be
considered as a direct numerical simulation of mesoscale effects.
For example, based on the authors knowledge, none
of these works considered the interaction of floated porosity
with dendrites. Furthermore, the mechanism of porosity nucleation,
in particular when regarding high quality melts, is still in question
which increases difficulties of mesoscale direct numerical
simulation (see, for example, chapter 7 of \citep{campbell2003c}). Since the focus of
this paper is on criterion function methods, it is worth to review
these methods in details.

Early work by \citet{pellini1953fwd} on introducing a thermal
criteria to evaluate the feeding ability reported that $G$ should be
greater than a critical value, $G_{cr}$, to avoid centerline
shrinkage in plate and bar castings. \citet{niyama1981psi} also
examined a number of commercial castings and confirmed that the
feeding gradient could be used to predict the formation of shrinkage
porosity. These researchers detected that, $G_{cr}$ depends on both
shape and size of a particular casting and could not be predicted in
advance. Later, \citet{niyama1982msp} discovered that both the scale
of porosity and the dendritic structure varied in proportion to the
size of the casting. As a result, they discovered that $G_{cr}$ is
proportional to $1/\sqrt{t_s}$. Therefore, their criterion function
can be expressed as $G\sqrt{t_s}$. Finally, they have approximated
$t_s$ by $(\theta_l - \theta_s)/R$ and presented $GR^{-1/2}$ as
their criterion function. The critical value of this criterion was
proven to be independent of the casting size, first by
\citet{niyama1982msp}, and later by other researchers on some
simple-shaped castings.

It is worth noting that replacing $\sqrt{t_s}$ by $1/\sqrt{R}$ stems
from an asymptotic analysis by
\citet{niyama1982msp} to justify their measure theoretically. This
analysis includes the solution of a one-dimensional quasi
steady-state Darcy flow within the mushy zone to compute the
pressure drop due to the friction of dendritic structure. The
current authors believe that this analysis not only does not
support the validity of the Niyama criterion but is also
misleading about the ability of this measure to predict
microporosity formation. In particular, the computation of the
Niyama criterion in the form of $GR^{-1/2}$ does not always give
reasonable results and it is recommended to use
$G\sqrt{t_s}$ instead \citep{carlson2002dnf,ou2002dnf,ou2005far,carlson2005fhn}.%
\footnote{The commercial package
\href{http://www.magmasoft.de/}{MAGMASOFT} also uses $G\sqrt{t_f}$
as the Niyama criterion} We shall address this issue in more
detail in \autoref{sec:analysisofcfm}.

On the other hand, the Niyama criterion is commonly used in real-world
designs, in particular in the case of steel castings. Consequently,
almost all commercial casting simulation packages have the ability to
use this measure. In research, the Niyama criterion has recently also
been used extensively by Carlson, Beckermann and coworkers to develop
new feeding-distance rules for low alloy
\citep{carlson2002dnf,ou2002dnf} and high alloy
\citep{ou2005far,carlson2005fhn} steels.

Although the Niyama criterion is more universal than the Pellini
criterion and resolves the size dependence problem, the shape
dependency of the critical value seems to be an unresolved issue.
The problem is due to this fact that (based on Pellini's
observation) the minimum temperature gradient required to feed a bar
is about 5--10 times greater than what is needed to feed a plate. As
mentioned in \citep{hansen1988mas}, the Niyama criterion fails to
solve the shape-dependency problem. \citet{hansen1988mas} did a
series of numerical simulations for bar and plate geometries with
different scales, i.e., multiplying all dimensions by a scaling
factor $N$. They showed that when the casting dimensions are scaled
by factor $N$, the temperature gradient, cooling rate and feeding
flow velocity are scaled by factors $N^{-1}$, $N^{-2}$ and $N^{-1}$,
respectively. The feeding flow velocity, $U$, was computed based on
the solution of mesoscale fluid flow governed by Darcy's law. Based
on these results, they try to remove the scale-dependency from
criterion functions that depend on $G$, $R$ and $U$ using a power
scaling; this leads to the criterion $G R^{-1/4} U^{-1/2}$. Although
their analysis just confirms that this criterion could be
scale-independent, they claimed the shape independency of this
measure too, i.e., the critical values of this criterion for plate
and bar geometries are the same. However, this claim is only
supported by their numerical experiment on plate and bar castings of
Pellini. To the best of the authors' knowledge, no further
experimental or numerical work supports this assertion.
Consequently, we believe that the analysis is not sufficient to
justify the shape-independency of this measure. In fact, we shall
argue in \autoref{sec:analysisofcfm} that  this measure cannot be
shape-independent as the shape dependency has a global nature which
cannot be summarized into local parameters such as $G$, $R$ and $U$.
The source of the shape-dependency of the Niyama criterion will be
discussed in \autoref{sec:analysisofcfm} as well. Since we have to
solve the flow equations coupled with energy equation to evaluate
Hansen and Sahm's criterion, it is computationally inefficient in
contrast to the Niyama criterion. It is worth noting that the Niyama
criterion is also scale-independent (for more details see
\citep{hansen1993sau,stefanescu2005css}).

\citet{sigworth1993mpf} also considered the formation of centerline
shrinkage in plate castings. They have proposed a geometric model in
which $G_{cr}$ is a function of the angle of the inner feeding
channel inside the casting, $\Theta_c$. Their criterion has the
following generic form: $G_x \sqrt{t_s} > c \ \tan{\Theta_c}$, where
$G_x$ is the temperature gradient along the major axis of the plate
and $c$ is a constant. These authors estimate a critical angle for
steel plates between 2 and 5 degree. In contrast to the previously
mentioned works which are constructed based on experimental
observations, this criterion resulted from a theoretical analysis.
Therefore, its limitation can be predicted based on the considered
assumptions. For example, it is obvious that this measure is not
universal and its application is limited to plate-like castings.
Further, applying this measure does not seems to be feasible in the
case of long freezing range alloys. The most important conclusion,
however, is that this alternative derivation results in the same
functional dependence between the temperature gradient and the
solidification time as that found experimentally by
\citet{niyama1982msp}. It turns out that the critical value of the
Niyama criterion should have a shape dependent component which
confirms the shape-dependency of this measure.

Recently, a dimensionless version of the Niyama criterion is
recently suggested by \citet{carlson2009prediction}. In this work it
was claimed that their criterion function could predict position and
quantity of defects without using a-priori known threshold value.
This claim is recently rejected in \cite{sigworth2009discussion}
(see also response letter \cite{carlson2009authors}). We believe
that our analysis in this paper partly support the discussion of
\citet{sigworth2009discussion} and hopefully resolve this challenge
by more rigorous reasoning.

Regarding long freezing range alloys, a criterion function of the form
$G t_f^{2/3}v_s^{-1}$ was
suggested by  \citet{lee1990mfb}. This criterion is supported
by combining theoretical analysis and experimental observations.
Following the scale-dependency study suggested in
\citep{hansen1988mas,stefanescu2005css}, it is obvious that this
criterion is neither scale- nor shape-independent.
According to \citep{lee2001mma,stefanescu2005css}
criterion function methods have so far had little success in predicting
porosity formation in the case of long freezing range alloys.

Prior to closing this section we would like to point out to an
elegant discussion on the shortcomings of the Niyama criterion by
\citet{spittle1994niyama} which is partly in agreement with our
discussion.

\section{Analysis of criterion function methods. }%
\label{sec:analysisofcfm}%

In this section we will analyze criterion function methods,
in particular regarding their limitations. To this end,
let $\Omega = \Omega_c \cup \Omega_m$  be the spatial domain, where $\Omega_c \subset \mathbb{R}^d$ ($d=$  2 or 3) is a
portion of $\Omega$ which includes the casting and $\Omega_m \subset
\mathbb{R}^d$ denotes the mold region. The interface between the
two is denoted by $\Gamma_{c-m} = \overline\Omega_c \cap
\overline\Omega_m$. From a macroscopic point of view, if the effect of melt
flow during solidification is neglected, solidification is
governed by heat conduction \citep{lewis2000fes,tavakoli2007usf}:%
\begin{equation}%
\label{heateqc}%
\rho_c c_c\ \frac{\partial\theta_c}{\partial t}= \nabla\cdot\ (k_c
\nabla \theta_c) + \rho_c L \frac{\partial f_s}{\partial t} \quad
\quad \textrm{in} \ Q_c = \Omega_c \times [0, T]
\end{equation}%
where $\rho_c$ is the cast density, $c_c$ is the cast specific heat,
$\theta_c$ is the temperature field in the casting region, $t$ is
the time variable, $k_c$ is the cast thermal conductivity, $L$ is
the fusion latent heat, $f_s$ is the solid fraction field, and $[0,
T]$ denotes the temporal domain. The initial condition is $\theta_c
= \theta_c^0 \quad \textrm{in}\ \Omega_c\times \{t=0\}$, where
$\theta_c^0$ is the pouring
temperature. In the mold region we have: %
\begin{equation}%
\label{heateqm}%
\rho_m c_m\ \frac{\partial\theta_m}{\partial t}= \nabla\cdot\ (k_m
\nabla \theta_m) \quad \quad \textrm{in} \ Q_m = \Omega_m \times [0,
T]
\end{equation}%
where $\rho_m,c_m,\theta_m,k_m$ are mold analogues of the variables
defined above. The initial condition in the mold
region is $\theta_m = \theta_\infty \quad \textrm{in}\
\Omega_m\times \{t=0\}$, where $\theta_\infty$ is the ambient
temperature. Natural boundary conditions are applied at the
mold-air interface, i.e., $k \frac{\partial\theta_m}{\partial
n_{m-a}} = h_{\infty} (\theta_{\infty}-\theta_m)$ on $\Gamma_{m-a}
\times [0, T]$, where $h_{\infty}$ denotes the mold-air convective
heat transfer coefficient, $\Gamma_{m-a}$ is the mold-air interface
and $n_{m-a}$ is the unit normal vector directed toward the
environment. The cast-mold
interface is modeled by the
interfacial heat transfer coefficient method \citep{manzari2000oht,tavakoli2007efd}, i.e.,%
$- k_c \frac{\partial\theta_c}{\partial n_i} = k_m
\frac{\partial\theta_m}{\partial n_i} = h_i (\theta_c - \theta_m)$
on $\Gamma_{c-m} \times [0, T]$, %
where $n_i$ is the unit normal vector directed toward the mold
and $h_i$ is the local heat transfer
coefficient of the cast-mold interface.

From a macroscopic point of view, if the effect of gravity on the
distribution of porosities is neglected, shrinkage porosities will
be located at the last solidified points, i.e., hot spots. In
particular, this holds for centerline shrinkages. Consequently, we
need to look for hot spots in a casting to predict the locations of
the centerline shrinkages. If we assume that the temperature field
is sufficiently smooth, the hot spots are a subset of the critical
points of the temperature field \citep{tavakoli2007ordb}. Since
shrinkage defects form near the freezing time, it is sufficient to
only consider critical points at $t=t_f$ or $\theta=\theta_s$, i.e.
we only consider the admissible set
$$\mathcal{A}_{ad} = \{\textbf{x}\in \Omega_c \mid \ |\nabla \theta
(\textbf{x}, t=t_f)|=0\},$$ or a suitable numerical approximation of
it. Note that this set contains not only hot spots, but also cold
spots and saddle points. A sufficient condition for  $\textbf{x}^*
\in {\cal A}_{ad}$ to be a hot spot is if the Hessian matrix of the
temperature field, $\textbf{H}_{\theta}(\textbf{x})=\nabla^2
\theta(\textbf{x},t=t_f)$, is negative definite at $\textbf{x}^*$,
i.e., for all vectors
$\textbf{d} \in \mathbb{R}^d, \textbf{d}\neq 0$,%
\begin{equation}%
\label{sufficientcondition}%
    \textbf{d}^T \textbf{H}_{\theta}(\textbf{x}^*) \ \textbf{d} < 0.%
\end{equation}%
An equivalent condition is if all eigenvalues of the Hessian are
negative. Later, we shall present a physical counterpart of
this sufficient condition.

As already mentioned in \autoref{sec:relatedworks},
\citet{niyama1981psi} observed that the scaling factor
$1 \sqrt{t_s}$ resolves the size-dependency problem of the Pellini
criterion. The authors justified their criterion through
one-dimensional quasi steady-state fluid flow within the dendritic
structure governed by the Darcy law. Their analysis showed that the
pressure drop due to the friction of dendritic structure is
inversely proportional to factor $GR^{-1/2}$ and they introduce
their measure as $GR^{-1/2}$ instead of $G \sqrt{t_s}$.

On the other hand, to the authors's knowledge, neither Niyama et
al.~nor other researchers provided experimental support for the
criterion $GR^{-1/2}$ as an interdendritic microshrinkage predictor
measure in a {\it{purely unidirectional}} solidifications. In fact,
all experimental support for the Niyama criterion are actually
related to  {\it{multi-directional}} solidifications, a concern
already mentioned in \cite{sigworth1993mpf,spittle1994niyama}. In
particular, \citet{spittle1994niyama} emphasized that available
supportive experimental data for the Niyama criterion are more
related to simple-shaped castings. Furthermore, in the case of
unidirectional solidification, the formation of shrinkage porosity
is depends on gravity \citep{kim1998pbs}, a factor not covered by
Niyama's analysis. Detailed discussion on this effect can be found
in \cite{tavakoli2009th}.

Moreover, it is important to note that the Darcy-law based
assumption of flow which is used in
\cite{niyama1981psi,carlson2009prediction} does not cover the whole
physics behind the defect formation. For example formation of defect
in the either form of void shrinkage or gas bubble needs defect
nucleation which is very important in small scales (cf.
\cite{campbell2003c,tavakoli2009th}). The nucleation phase is so
important that it can completely change the distribution of defects.
Due to importance of this issue we briefly notify this effect here
(more details can be found in \cite{tavakoli2009th}).

In general formation of void in a clean melt is very difficult as
tensile strength of liquid metals is very large (cf. \cite[ch. 7 of
][]{campbell2003c}). According to \cite[ch. 7 of ][]{campbell2003c},
homogeneous nucleation of void region is almost impossible. And
formation of void regions are aided either by formation of gas
porosity or nucleation (in fact initiation) on the pre-existing
bubbles in the melt (which are not available for high quality
melts). In the case of clean melts, if shape of casting be
appropriate (e.g. plate-like castings), shrinkage can be compensated
by surface sink and no internal shrinkage (cf. Figure 7.4 of
\cite{campbell2003c}). This theory is also supported by experimental
works in \cite{campbell1969opl,awano2004sms,tavakoli2009th}. In the
appendix section of this paper we briefly recall results of our
experimental works in this regard.

Moreover, to study Darcy law based micro-defect formation, it is
required to consider a two-phase flow in porous structure formed by
dendrites (gas or void as dissolved and/or distinct phase in
addition to the melt). Moreover, the gravity has an important effect
on the flow which is easily ignored by the mentioned authors. To see
effect of gravity, interested peoples are refereed to the following
experimental works
\cite{sulfredge1990svf,tagavi1990vfu,sulfredge1992vfr,sulfredge1993via,han2006mbm}

After this brief discussion, we belive that the Niyama's criterion
gives its feasibility according to a macro-scale analysis. Here we
justify it using the Chvorinov's rule instead of the simplified
Darcy law. According to \citep{flemings1974sp}, the following
relation can be obtained by a unidirectional analysis of metal
solidification in a low conductivity mold:%
\begin{equation}%
\label{onedanalplate}%
    M_c = \frac{V_c}{A_c} = \frac{2}{\sqrt{\pi}}%
    \big(\frac{\theta_M-\theta_\infty}{\rho_c L}\big)%
    \sqrt{k_m \rho_m c_m}%
    \sqrt{t_s},%
\end{equation}%
where $M_c$ is the geometrical modulus of the casting, $V_c$ the
casting volume and $A_c$ the casting surface area trough which heat
is lost, and $\theta_M$ the melting point. Chvorinov's rule is
usually expressed as%
\begin{equation}%
\label{chvorinov}%
    M_c = \mathcal{C} \sqrt{t_s}%
\end{equation}%
with a constant $\mathcal{C}>0$. According to Chvorinov's rule the
square root of the freezing time is proportional to the casting
size. So it is possible to remove the size effect from the Pellini's
critical feeding gradient using term $\sqrt{t_s}$. This leads to
measure $G\sqrt{t_s}$ which is identical to Niyama et. al.
observations, but this time with a fairly feasible macro-scale heat
transfer analysis.

However, Chvorinov's rule is derived based on a unidirectional
solidification analysis, and consequently can not predict the total
solidification time of an arbitrary geometry. For example, in the
case of bar casting, the solidification condition coincides with
Chvorinov's assumptions in the early stage of solidification, but
later the mode of solidification change from unidirectional to
multidimensional and the cooling rate is considerably increased
\citep{santos1983atd,santos1998tbd}. According to
\citet{flemings1974sp}, the generalized counterpart of
\autoref{onedanalplate} for cylindrical and spherical geometries
has the following form%
\begin{equation}%
\label{multidanalch}%
    M_c = \frac{V_c}{A_c} = %
    \frac{\theta_M-\theta_\infty}{\rho_c L}%
    \big( %
    \frac{2}{\sqrt{\pi}} \sqrt{k_m \rho_m c_m}\sqrt{t_s}%
    + \frac{n k_m t_s}{2r}%
    \big),%
\end{equation}%
where $n = 0, 1$ and $2$ for plate, cylinder and sphere,
respectively, and $r$ is the radius of the casting. Consequently, a
generalized Chvorinov's rule has the following form%
\begin{equation}%
\label{mchvorinov}%
    M_c = \mathcal{C} \big( \sqrt{t_s} + \frac{\sqrt{\pi \alpha_m}}{4} \frac{n}{r}\ t_s\big)%
\end{equation}%
where $\alpha_m = k_m/\rho_m c_m$ is the mold thermal diffusivity.
Comparing \autoref{chvorinov} with \autoref{mchvorinov}, it is clear
that the simple Chvorinov and Niyama
approximations are valid  if the mold thermal diffusivity
or the overall mean curvature of the casting, $n/r$, are small.
This suggests that a generalized Niyama criterion would have
the form%
\begin{equation}%
\label{mniyama}%
    G \big( \sqrt{t_s} + \frac{\sqrt{\pi \alpha_m}}{4} \frac{n}{r}\ t_s\big) > \mathcal{B}%
\end{equation}%
with a constant $\mathcal{B}$ that depends on the physical properties
of mold and metal as well as the ambient temperature. To determine
$n$ and $r$ in \autoref{mniyama}, some geometric reasoning
procedures are required. Of course, this formula appears to be in
conflict with
\citet{niyama1981psi} as that work was on cylindrical geometries not
plate-like ones. The discrepancy is easily resolved by
computing the relation of the two terms in
parentheses of \autoref{mniyama} for the cases given in \citet{niyama1981psi}:
using physical properties and casting conditions
given in \citep{niyama1981psi}, the relative size of the second term
is smaller than $0.01$ for the all test cases.

According to the above discussion, we would like to accept the
Niyama criterion as a macroscopic shrinkage prediction measure as it
only detect regions around hot centers. We are hopeful that our
theoretical justification remove a-priori made miss-leading about
which is due to Niyama analysis based on micro fluid flow within
mushy zone. Therefore, we can not confirm the claim of
\cite{carlson2009authors} which is already mentioned in this paper.
Moreover, our analysis show that the shallow thermal gradient
regions, the Niyama criterion detects are not essentially hot spots.
But they can be also cold spots or saddle points. Such points
usually happen in complicated castings not simple plate-like ones
examined in literature. To cope this limitation, it is possible to
check for second order derivative of the temperature field. It is
worth to mention that almost all of the experimental support for the
Niyama criterion are related to simple shaped castings and detected
defects regions are related to nothing else regions around hot-spots
(as complement of discussion here see also appendix).

Moreover, it is reasonable to consider the thermal criterion
function as macroscopic defect prediction measures (in fact they
detects nothing else hot-spots). Therefore, regarding to long
freezing range alloys for which defects are not concentrated at hot
centers, we are unable to simply use thermal criteria. Therefore,
the focus of this study is on the short freezing range
solidifications in which the solid/liquid interface can be
considered as a macroscopically planar interface. As it is
experimentally shown in \cite{awano2004sms}, increasing the
concentration of dissolved gas in the melt changes the distribution
of defects from concentrated macro-shrinkages to distributed
micro-porosities. This is due to ease of defect (in connection to
gas porosity) nucleation near solid/liquid interface during
solidification (cf. alos Figure 7.26 of \cite{campbell2003c}).
Therefore, we also have to limit application of thermal criteria to
high quality melts which are sufficiently degassed too.

As a final remark in this section we note that the generalized
Niyama criterion introduced here is expected to be less dependent on
the shape of a casting. This is due to the contribution of the extra
term that models the casting's overall mean curvature, $\kappa=\frac
nr$. However, the overall mean curvature is a poor measure in the
case of complicated geometries. Instead, one may use a criterion
that considers $\kappa$ to be a local quantity $\kappa(\textbf{x})$
that depends on $\textbf{x}$: it would denote an effective curvature
at a point $\textbf{x}$ by only considering the geometry within a
neighborhood of diameter $\cal R$ around $\textbf{x}$, where $\cal
R$ is smaller than the diameter of the casting but larger than the
maximum distance of points in the cast from the mold in the
neighborhood of the desired point. We will not directly pursue this
line of thought further, however, and leave it for future research.
However, we shall consider the role of local mean curvature in
\autoref{sec:newmethod} where we introduce our criterion function.

\section{New thermal criterion function. }%
\label{sec:newmethod}%

\begin{figure}[ht]%
\begin{center}%
\ifshowfigs%
\includegraphics[width=12.cm]{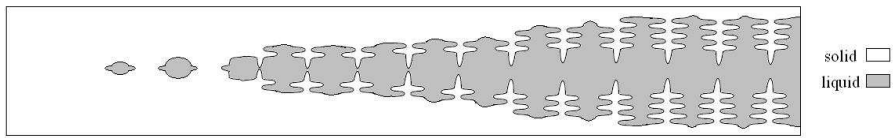}%
\fi%
\caption{Formation of a centerline shrinkage and some dispersed mesoscopic shrinkages along its major axis.}%
\label{fig:mesocenterlineshr}%
\end{center}%
\end{figure}

The number, size and distribution of shrinkage defects depend on the
physical properties of the melt and mold, quality of the melt and
the geometry of casting. If the effect of gravity on the
distribution of shrinkage defects is neglected, the last
solidification points are located at the spatial position of local
hot spots. Therefore, to predict the location of macroscopic
shrinkages, it is sufficient to determine the position of local hot
spots (see \autoref{sec:analysisofcfm}). As shown schematically in
\autoref{fig:mesocenterlineshr}, mesoscale shrinkages are usually
formed during the evolution of a solidifying front to form a
macroscopic shrinkage. In fact, an elongated liquid pool usually
leaves some dispersed mesoscale shrinkages along its major axis. It
is obvious that the width of the mushy zone or sizes of dendrites
affect on the distribution and morphology of these mesoscale
shrinkages. We have exaggerated this effect in the figure for the
purpose of exposition.

\begin{figure}[ht]%
\begin{center}%
\ifshowfigs%
\includegraphics[width=14.cm]{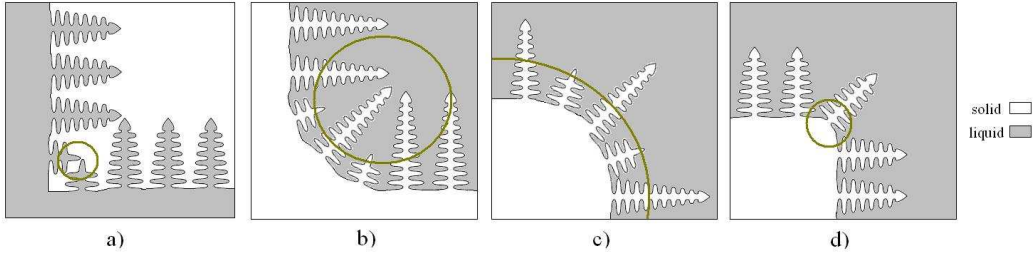}%
\fi%
\caption{Effect of the macroscopic interface curvature on the
formation of shrinkage defects: Increasing interface curvature from
a to d; a and b show negative curvature (concave interface) c and d
show positive curvature (convex interface). Circles
show underlying macroscopic curvature.}%
\label{fig:interfacecurvatures}%
\end{center}%
\end{figure}%

In order to construct an indicator function for the formation of
centerline shrinkages and related dispersed mesoscale shrinkage
around their wakes the (signed) local mean curvature
\citep{giga2006see} of the macroscopic solid/liquid interface,
$\kappa= - \nabla \cdot \hat{\textbf{n}}$, is a good measure to
predict the closeness of the solid/liquid interface. We use a
critical value, $\kappa_{cr}$, to predict the formation of shrinkage
defects. Obviously, the magnitude of $\kappa_{cr}$ should decrease
as the width of the mushy zone increases; it will also depend on the
size and distribution of crystals as well as local solidification
conditions cf. \citep[Figure 4.18]{kurz1998fos} and other physical
parameters. Notice that when the local mean curvature of the
solid/liquid interface is positive, i.e., when the interface is
convex, the danger of shrinkage formation is minimal (see
\autoref{fig:interfacecurvatures}). We therefore only consider the
case $\kappa_{cr}<0$, and assume the following for for our
macroscopic criterion function:%
\begin{equation}%
\label{macromeasure}%
    \mathcal{F}_\Sigma (\kappa) = a + b \kappa > \kappa_{cr} = \kappa_{cr}(\mathcal{M}, G, R, v_s),
\end{equation}%
where $\mathcal{F}_\Sigma$ denotes the macroscopic defect predictor
measure, $a,b$ are constants and $\mathcal{M}$ denotes the
contribution of materials properties on $\kappa_{cr}$. Models
including higher powers of $\kappa$ are obvious extensions. Let us
note that the sign of the local curvature of the temperature field
$\kappa$ corresponds to the local convexity of the temperature
Hessian \citep{kobayashi1963fdg}: when the local curvature is
positive the Hessian matrix is locally positive definite and vice
versa. In other words, the critical interface curvature criterion
above covers the required sufficient and necessary conditions
mentioned in \autoref{sec:analysisofcfm}. On the other hand,
criteria like the one above can not predict the size of
macroshrinkages. However, we are hopeful to predict the spatial
extent of macroshrinkages by predicting the dispersed meso-scale
defects in the vicinity of each macroshrinkage (such dispersed
defects are also causes this miss-leading that thermal criteria are
micro-shrinkage prediction measures).

Now, let us to extend \autoref{macromeasure} by incorporating the
dependency of the critical value to the field solution. For this
purpose, we add meso-scale information to the suggested macroscopic
measure:

\noindent {\it\bf Model I: Using the mushy zone width. } The width
$L$ of the mushy zone
can be estimated by%
\begin{equation}%
\label{mushywidth}%
    L \approx \frac{\theta_l-\theta_s}{G}.
\end{equation}%
Since with increasing the mushy zone width the absolute value of
$\kappa_{cr}$ in \autoref{macromeasure} should decrease, the
following measure can be introduced:%
\begin{equation}%
\label{macromesomeasure1}%
    \mathcal{F}_{\Sigma\xi} (\kappa, G) = \big( a + b \kappa \big) L^c > \kappa_{cr} = \kappa_{cr}(\mathcal{M}),%
\end{equation}%
where $\mathcal{F}_{\Sigma\xi}$ denotes the macroscopic criterion
which is enriched with mesoscale information, and $c$ is a
constant. For our numerical studies below, we will choose $a=0$ and all of the
other constants equal to one, yielding
\begin{equation}%
\label{macromesomeasureI}%
    \mathcal{F}_{\Sigma\xi}^I (\kappa, G) = \kappa G^{-1} > \kappa_{cr} = \kappa_{cr}(\mathcal{M})%
\end{equation}%

\noindent {\it\bf Model II: Using the primary dendritic arm spacing.
} According to \citet{kurz1998fos}, the primary dendritic arm
spacing, $D$, or crystal size, can be approximated by the following relation,%
\begin{equation}%
\label{DAS}%
    D \approx \mathcal{K}\ v_s^{-1/4} G^{-1/2},%
\end{equation}
where $\mathcal{K}$ denotes a material constant. We expect the
absolute value of $\kappa_{cr}$ to decrease with increasing
dendritic arm spacing,
resulting in%
\begin{equation}%
\label{macromesomeasure2}%
    \mathcal{F}_{\Sigma\xi} (\kappa, D) = \big( a + b \kappa \big) D^c > \kappa_{cr} = \kappa_{cr}(\mathcal{M}).%
\end{equation}%
With the same choice of constants, we get%
\begin{equation}%
\label{macromesomeasureII}%
    \mathcal{F}_{\Sigma\xi}^{II} (\kappa, v_s, G) = \kappa v_s^{-1/4} G^{-1/2} > \kappa_{cr} = \kappa_{cr}(\mathcal{M})%
\end{equation}%

\noindent {\it\bf Model III: A heuristic reasoning. } It is clear
that by increasing speed of solid/liquid interface the time for the
mass diffusion is decreased. Moreover, for a constant front speed,
feeding ability is increased by increasing temperature gradient in
the vicinity of interface. As a result parameter $v_s^c/G^d$ could
shows difficulty of melt feeding. Assuming constants $c$ and $d$ are
equal to one, the following heuristic criterion can be introduced
\begin{equation}%
\label{macromesomeasureIII}%
    \mathcal{F}_{\Sigma\xi}^{III} (\kappa, v_s, G) = \kappa v_s G^{-1} > \kappa_{cr} = \kappa_{cr}(\mathcal{M})%
\end{equation}%

\noindent {\it\bf Model IV: Generalized model. } The above results
suggest that realistic criterion functions should depend on
$\kappa$, $v_s$ and $G$. A criterion generalizing all of the above
relationships could then have the following form:%
\begin{equation}%
\label{macromesomeasureIV}%
    \mathcal{F}_{\Sigma\xi}^{IV} (\kappa, v_s, G) = \big( a + b \kappa \big) v_s^c G^d > \kappa_{cr} = \kappa_{cr}(\mathcal{M})%
\end{equation}%
Determining the unknown constants $a, b, c, d$ for a specific alloy and
casting conditions, quantitative experimental data are obviously required.
They can then be computed using parameter
estimation procedures such as the least-squares method \citep{nocedal1999no}.

\section{Results and discussion. }
\label{sec:result}%

In this section will evaluate the presented method on some known
test cases for which experimental results are available. The energy
equation is solved by the classical explicit finite difference
method and the effect of phase change is incorporated into the heat
equation using the effective heat capacity method
\citep{hong2004cmh}. A personal computer with an AMD Athlon
2.41 Ghz CPU and 3GB RAM
was used as the computing platform in this study.%

\subsection{Pellini's plate and bar casting. }%
\label{sec:pellinitest}%

In this section the, we consider sand mold casting of plain carbon
steel. We choose two rectangular geometries with $W/T = 1$ (bar) and
$W/T= 3$ (plate), respectively; in each case, L/T = 9, where $T$,
$W$ and $L$ denote the casting thickness, width and length.
Following Pellini's experimental data \citep{pellini1953fwd}, the
casting thickness is taken to be equal to 2 and 4 inches
(approximately 5 and 10 cm) for both geometries. Physical properties
and initial and boundary conditions used here presented in
\autoref{materprop_pellini}. According to \citet{pellini1953fwd}, in
2 inch-thick plates, shrinkage porosity occurred in an area where
the feeding gradient ($G^+$ in this case) was less than 20--40
$^o$C/cm. Regarding to 4-inch bars a higher feeding gradient is
required to prevent the centerline shrinkage: 120--240 $^o$C/cm. We
discretize the domain by a three-dimensional uniform Cartesian grid with mesh
size $T/30$ in this study.

\begin{table}[ht]%
\caption{Physical properties, initial and boundary conditions used
in Pellini's plate and bar casting test cases. Units are in SI, except
for the temperature which is expressed in $^o$C.}%
\label{materprop_pellini}%
\centering %
\begin{tabular}{cccccccccc}%
\\\hline\hline%
&$k$&$\rho$&$c$&$L$&$\theta^0$&$\theta_l$&$\theta_s$&$h_i$&$h_\infty$\\\hline\hline\\%
metal&33.5&7200&627&2.7 $\times{10}^5$&1594&1490&1440&418&-\\%
sand&0.7&1500&1130&-&20&-&-&-&75\\ \hline\hline%
\end{tabular}%
\end{table}%

\begin{figure}[ht]%
\begin{center}%
\ifshowfigs%
\includegraphics[width=15.cm]{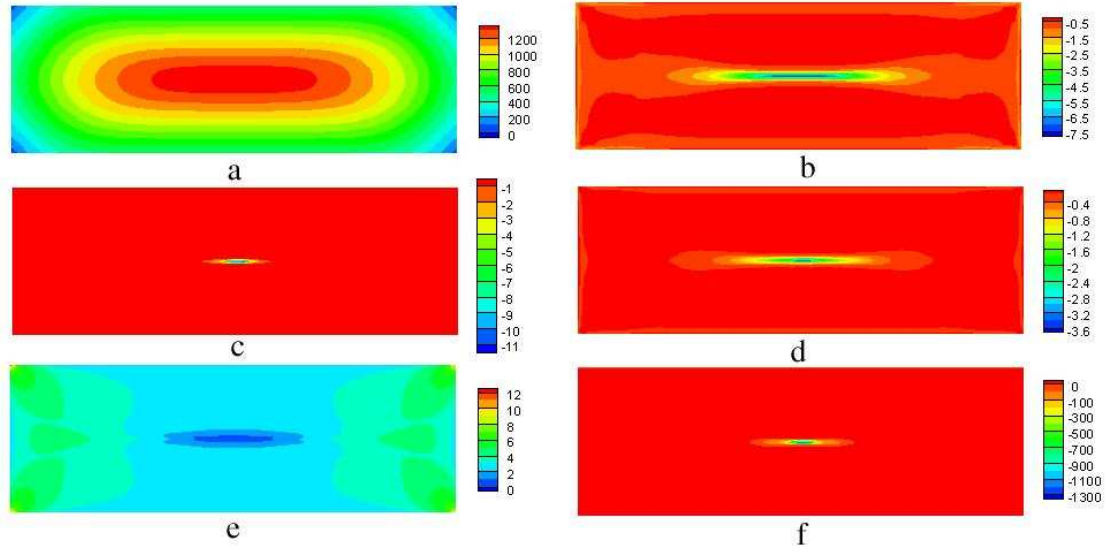}%
\fi%
\caption{Results of simulations for Pellini's plate casting, 2D cross section at T/2: %
a)~Local freezing time in seconds, %
b)~Interface curvature map in cm, %
c)~$\mathcal{F}_{\Sigma\xi}^{I}$ in deg$^{-1}$, %
d)~$\mathcal{F}_{\Sigma\xi}^{II}$ in deg$^{-1/2}$ min$^{1/4}$ cm$^{-3/4}$, %
e)~Niyama criterion in deg$^{1/2}$ min$^{1/2}$ cm$^{-1}$, %
f)~$\mathcal{F}_{\Sigma\xi}^{III}$ in deg$^{-1}$ min$^{-1}$ cm.}%
\label{fig:pellini-wrt3}%
\end{center}%
\end{figure}

\begin{figure}[ht]%
\begin{center}%
\ifshowfigs%
\includegraphics[width=15.cm]{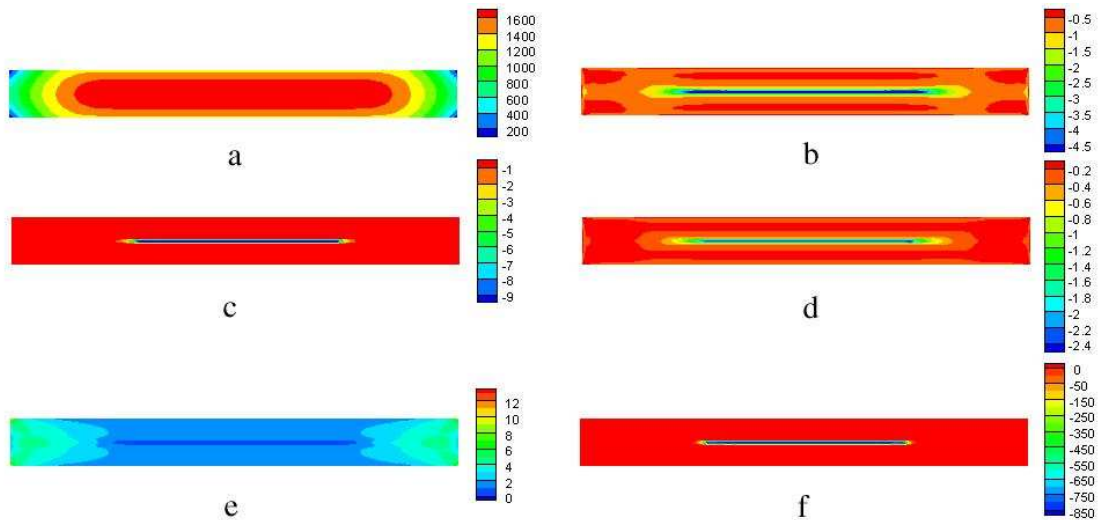}%
\fi%
\caption{Results of simulations for Pellini's bar casting, 2D cross section at T/2: %
a)~Local freezing time in seconds, %
b)~Interface curvature map in cm, %
c)~$\mathcal{F}_{\Sigma\xi}^{I}$ in deg$^{-1}$, %
d)~$\mathcal{F}_{\Sigma\xi}^{II}$ in deg$^{-1/2}$ min$^{1/4}$ cm$^{-3/4}$, %
e)~Niyama criterion in deg$^{1/2}$ min$^{1/2}$ cm$^{-1}$, %
f)~$\mathcal{F}_{\Sigma\xi}^{III}$ in deg$^{-1}$ min$^{-1}$ cm.}%
\label{fig:pellini-wrt1}%
\end{center}%
\end{figure}

\autoref{fig:pellini-wrt3} and \autoref{fig:pellini-wrt1} show the
results of this numerical experiment, in which contour plots of
the local freezing time, curvature map,
$\mathcal{F}_{\Sigma\xi}^{I}$, $\mathcal{F}_{\Sigma\xi}^{II}$,
$\mathcal{F}_{\Sigma\xi}^{III}$ and the Niyama criterion are
included. The plots show that regions with a large negative
curvature coincide with centerline shrinkage. Therefore, the
curvature value can be used (at least) as a qualitative defect
indicator measure. Further, the suggested criteria in this study
provide a sharp contrast between the centerline shrinkage and the
healthy parts of the castings. However, the contrast provided by the
third suggested measure, $\mathcal{F}_{\Sigma\xi}^{III}$, is
superior to the others.


\begin{figure}[ht]%
\begin{center}%
\ifshowfigs%
\includegraphics[width=15.cm]{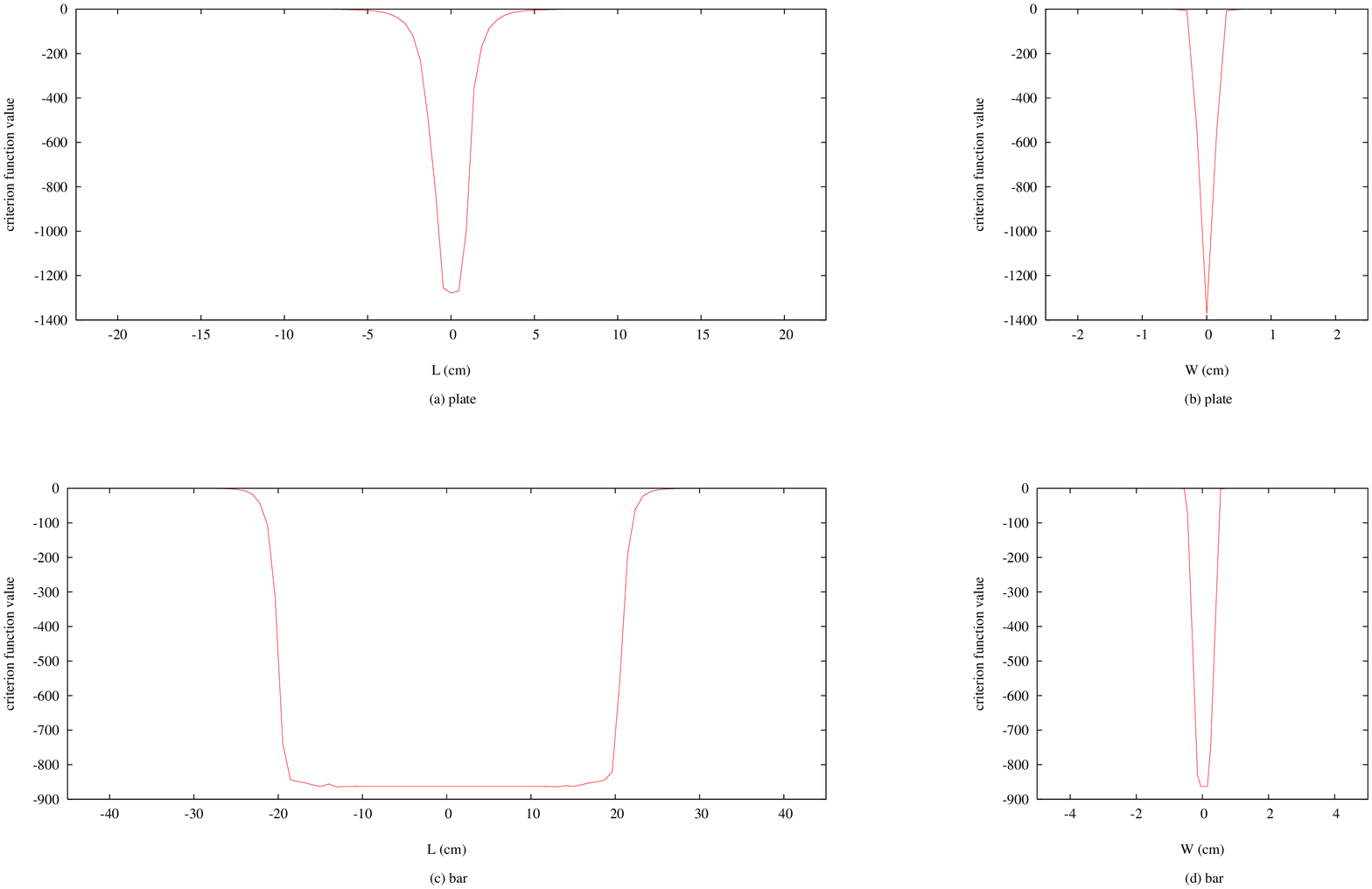}%
\fi%
\caption{Results of simulations for Pellini's test cases. Left:
Variation of $\mathcal{F}_{\Sigma\xi}^{III}$ along a horizontal line
through the casting center.
Right: Variation along a vertical line
through the casting
center. Top: Plate. Bottom: Bar.}%
\label{fig:linevarpb}%
\end{center}%
\end{figure}

\begin{figure}[ht]%
\begin{center}%
\ifshowfigs%
\includegraphics[width=15.cm]{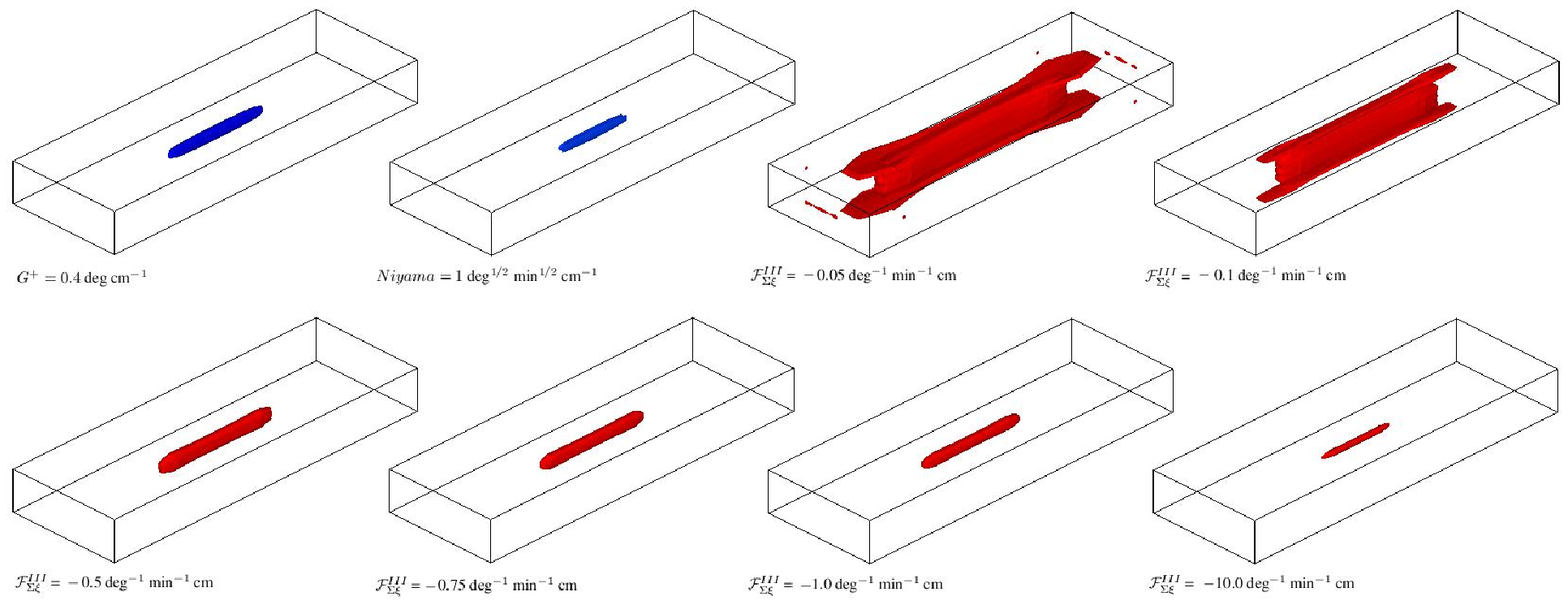}%
\fi%
\caption{Results of simulations for Pellini's plate casting:
Iso-contours of critical feeding gradient ($G^+$) and critical
Niyama value in conjunction with various iso-contours of
$\mathcal{F}_{\Sigma\xi}^{III}$.}%
\label{fig:pellini-ic3}%
\end{center}%
\end{figure}

\begin{figure}[ht]%
\begin{center}%
\ifshowfigs%
\includegraphics[width=15.cm]{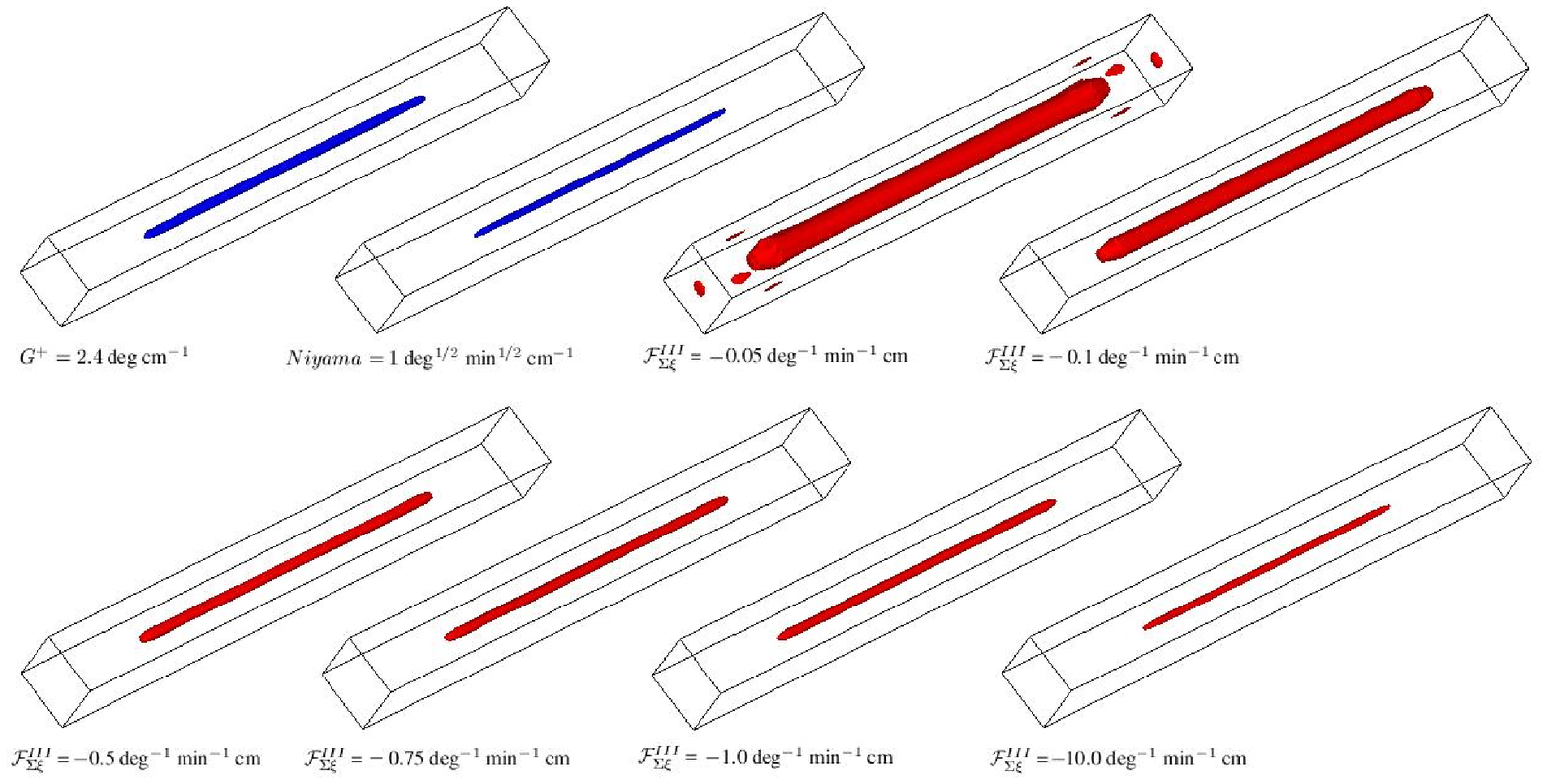}%
\fi%
\caption{Results of simulations for Pellini's bar casting:
Iso-contours of the critical feeding gradient ($G^+$) and critical
Niyama value in conjunction with various iso-contours of
$\mathcal{F}_{\Sigma\xi}^{III}$.}%
\label{fig:pellini-ic1}%
\end{center}%
\end{figure}

\autoref{fig:linevarpb} shows the variation of
$\mathcal{F}_{\Sigma\xi}^{III}$ along lines parallel to either the
width or length of the castings and through the center of the
castings. The plots show the desired sharp contrast of the criterion
value in the vicinity of the centerline shrinkage. This property
makes the selection of appropriate threshold values for this
criterion much simpler as the region where shrinkage is predicted to
occur becomes largely insensitive to the choice of threshold value.
To select a useful threshold, various iso-contours of
$\mathcal{F}_{\Sigma\xi}^{III}$ in conjunction with the
corresponding critical feeding gradient (based on Pellini's
measurements) and the critical Niyama value are plotted in
\autoref{fig:pellini-ic3} and \autoref{fig:pellini-ic1} for plate
and bar geometries, respectively. The plots show that the
iso-contours $\mathcal{F}_{\Sigma\xi}^{III}$ = -0.5 to -10.0
deg$^{-1}$ min$^{-1}$ cm seems to be a good range for a critical
value. Note that we do not want to make a sharp decision on the
critical value as we are still believed to the mentioned
shape-dependency problem (though our measure introduced to moderate
it). Moreover, our results show that critical iso-contours of Niyama
and Pellini criterion are not identical to each other. But they are
more qualitatively matched.


\subsection{Niyama's cylinders casting. }%
\label{sec:niyamatest}%

\citet{niyama1982msp} presented experimental results for cast
vertical cylinders of different diameters (3, 6 and 9 cm) with a top
riser and mold made from furan-banded silica sand. They had five
different steels ranging from low alloy to high alloy steels. They
showed that the critical temperature gradient, $G_{cr}$, which is
required to avoid shrinkage porosity is about 24, 12 and 8 deg
cm$^{-1}$ for 3, 6 and 9 cm diameter cylinders, respectively.
Further, they showed that the critical value for Niyama's
criterion is about 1 deg$^{1/2}$ min$^{1/2}$ cm$^{-1}$
independent from the casting size. Simulation parameters used in our
computations are given in \autoref{materprop_niyama}. Niyama's
cylinders are discretized by a uniform three-dimensional Cartesian grid with
mesh size $D/30$, where $D$ denotes the cylinder diameter.%

\begin{table}[ht]%
\caption{Physical properties, initial and boundary conditions used
in Niyama's cylinders casting test cases. Units are in SI, except
for the temperature which is expressed in $^o$C.}%
\label{materprop_niyama}%
\centering %
\begin{tabular}{cccccccccc}%
\\\hline\hline%
&$k$&$\rho$&$c$&$L$&$\theta^0$&$\theta_l$&$\theta_s$&$h_i$&$h_\infty$\\\hline\hline\\%
metal&33.5&7200&712&2.7 $\times{10}^5$&1610&1520&1485&infinite&-\\%
sand&0.84&1500&1130&-&20&-&-&-&21\\ \hline\hline%
\end{tabular}%
\end{table}%

\begin{figure}[ht]%
\begin{center}%
\ifshowfigs%
\includegraphics[width=15.cm]{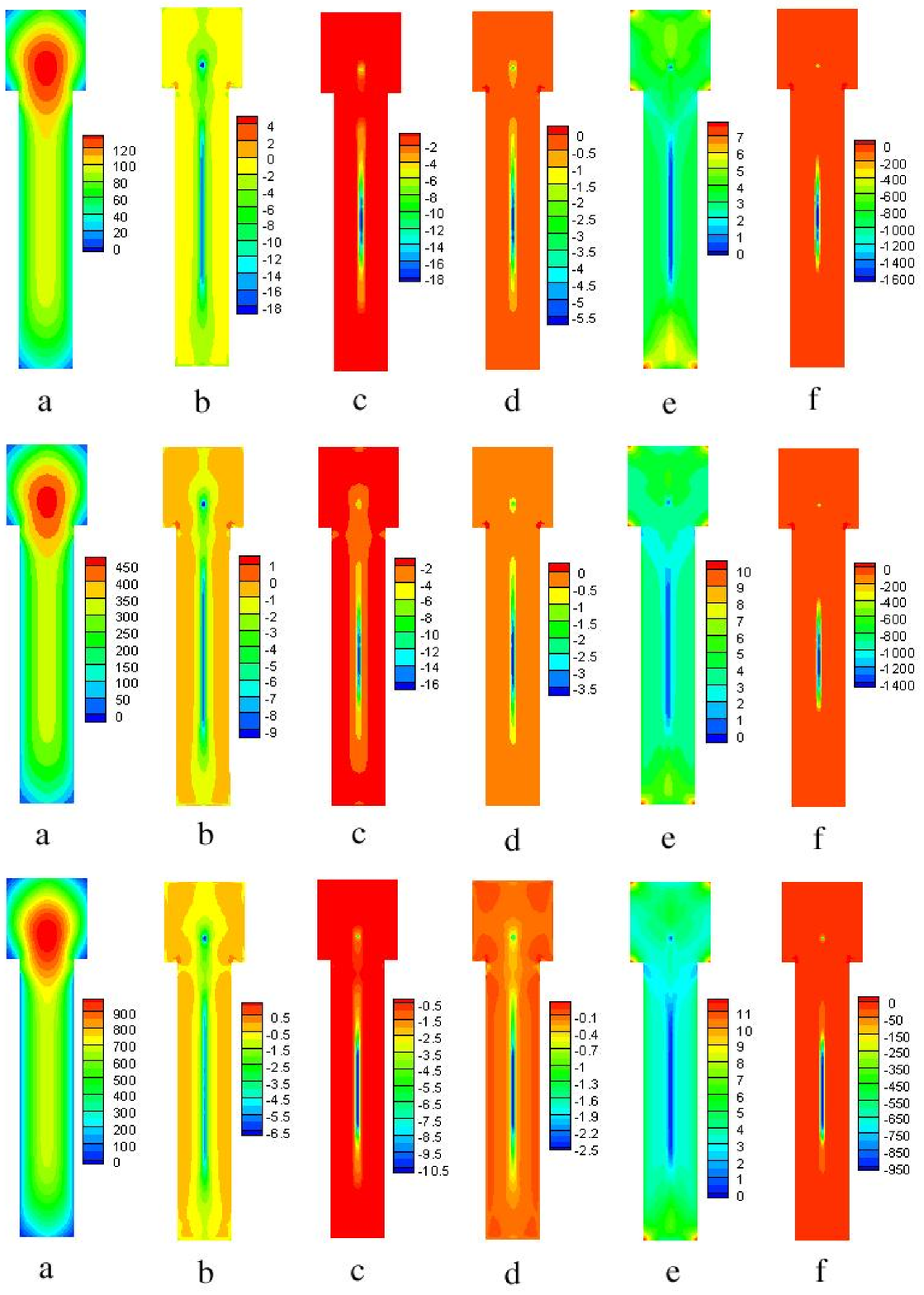}%
\fi%
\caption{Results of simulations for Niyama's cylinder casting, %
2D cross section at D/2: %
a)~Local freezing time in seconds, %
b)~Interface curvature map in cm, %
c)~$\mathcal{F}_{\Sigma\xi}^{I}$ in deg$^{-1}$, %
d)~$\mathcal{F}_{\Sigma\xi}^{II}$ in deg$^{-1/2}$ min$^{1/4}$ cm$^{-3/4}$, %
e)~Niyama criterion in deg$^{1/2}$ min$^{1/2}$ cm$^{-1}$, %
f)~$\mathcal{F}_{\Sigma\xi}^{III}$ in deg$^{-1}$ min$^{-1}$ cm. %
Top: $D = 3$ cm. %
Middle: $D = 6$ cm. %
Bottom: $D = 9$ cm.}%
\label{fig:niyama-2d}%
\end{center}%
\end{figure}

\autoref{fig:niyama-2d} shows the results of this numerical
experiment, in which the contour plot of the local freezing time,
curvature map, the three criteria introduced in this study and the
Niyama criterion are included. The validity of the curvature map as
a qualitative defect indicator measure is also obvious in this
numerical experiment. Further, the sharp contrast of the suggested
criteria to identify the centerline shrinkage, in particular
of $\mathcal{F}_{\Sigma\xi}^{III}$ is obvious.

\begin{figure}[ht]%
\begin{center}%
\ifshowfigs%
\includegraphics[width=15.cm]{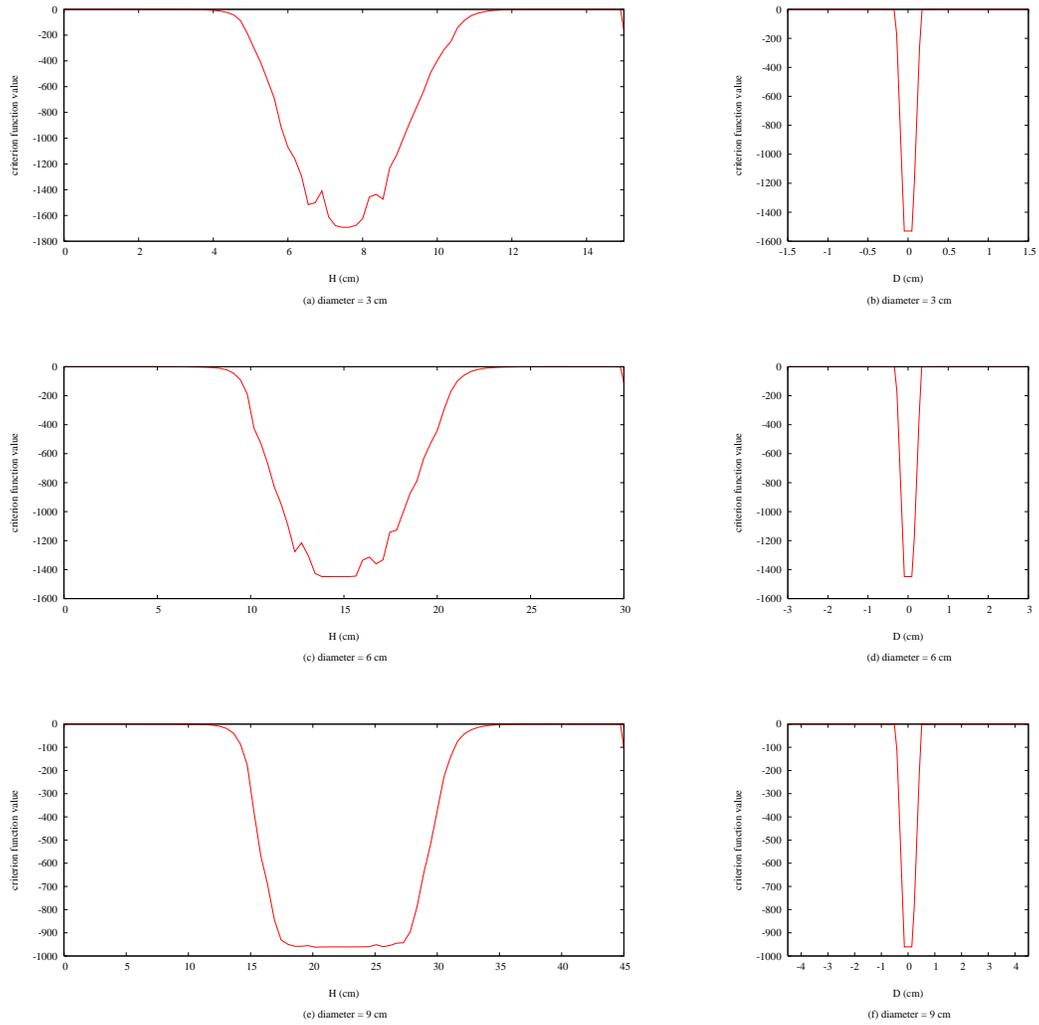}%
\fi%
\caption{Results of simulations for Niyama's cylinders casting. Left:
Variation of $\mathcal{F}_{\Sigma\xi}^{III}$ along the
objects centerline. Right:
Variation along a horizontal line through
the cylinder's center.
Top: $D = 3$ cm. Middle: $D = 6$ cm. Bottom: $D = 9$ cm.}%
\label{fig:linevarcyl}%
\end{center}%
\end{figure}

\begin{figure}[ht]%
\begin{center}%
\ifshowfigs%
\includegraphics[width=15.cm]{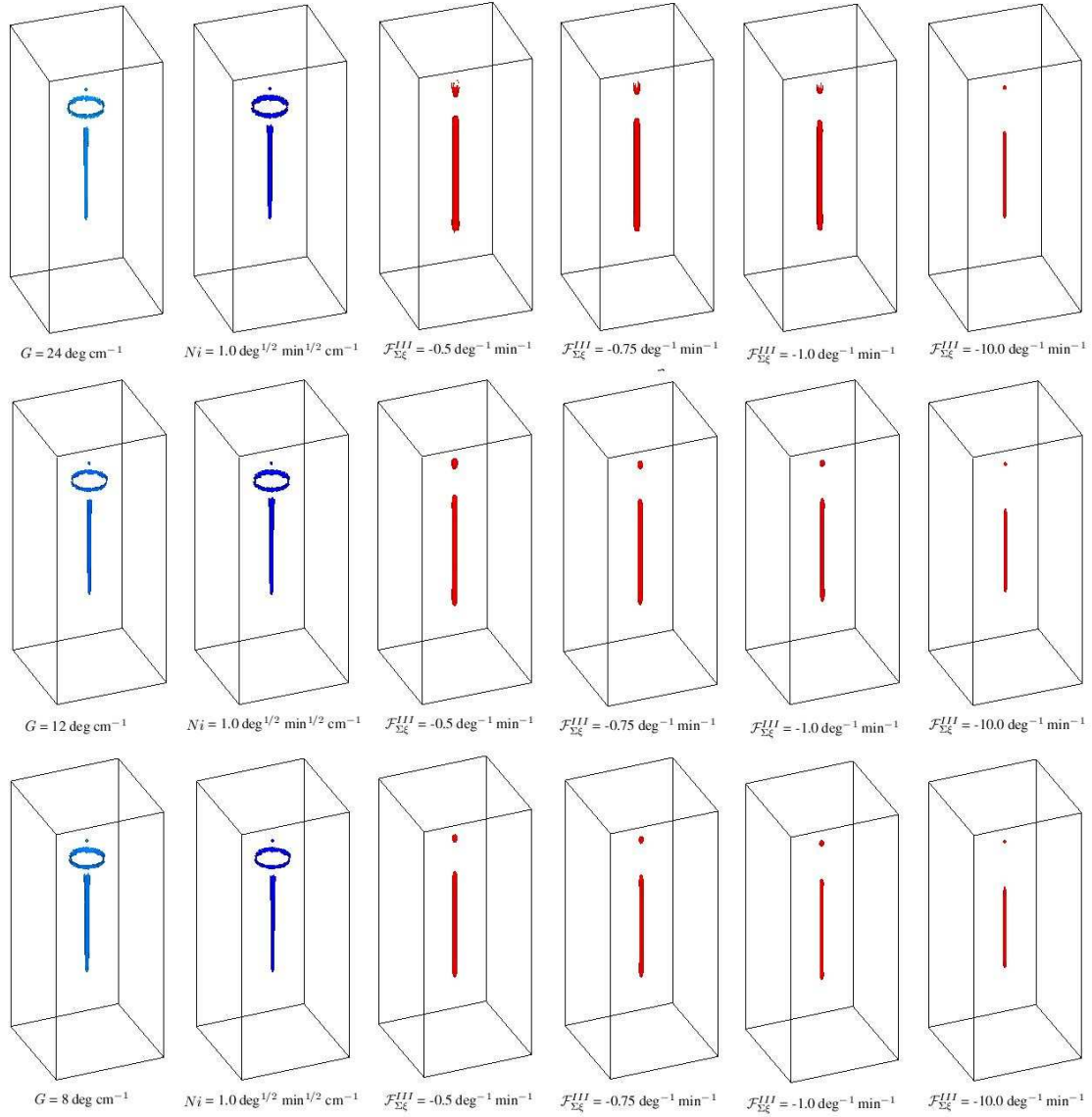}%
\fi%
\caption{Results of simulations for Niyama's cylinders casting:
iso-contours of the critical feeding gradient ($G$) and critical
Niyama value ($Ni$) in conjunction with various iso-contours of
$\mathcal{F}_{\Sigma\xi}^{III}$.}%
\label{fig:niyama-ic}%
\end{center}%
\end{figure}

\autoref{fig:linevarcyl} shows the variation of
$\mathcal{F}_{\Sigma\xi}^{III}$ along scan lines which are parallel
to either the diameter or height of the cylinders and through the
centers of the cylindrical parts of the geometry. This plot again
shows the sharp contrast of $\mathcal{F}_{\Sigma\xi}^{III}$ in the
vicinity of defects, making the choice of a threshold value a simple
one. To do so, various iso-contours together with the corresponding
critical feeding gradient (based on Niyama's results) and the
critical Niyama value are plotted in \autoref{fig:niyama-ic} for
Niyama's cylinders. Again, the iso-contour
$\mathcal{F}_{\Sigma\xi}^{III} = -0.75 to -10.0 $ deg$^{-1}$
min$^{-1}$ cm appears to be a good range for critical values.
However, we should emphasis that due to provided high contrast of
the suggested measure there is more freedom for selection of a
critical value (though in general there is no universal critical
value as measure is not essentially shape-independent).

These results of indicate that the suggested criterion and the
Niyama or Pellini criteria predict different shapes for the defect
region, as shown by the presence or absence of the ring-like shape
around the riser in the vicinity of riser-cast connection. We
believe that this discrepancy results from the fact that the latter
criteria only use the sufficient condition mentioned in
\autoref{sec:analysisofcfm} and violate the necessary condition,
whereas our criterion properly takes it into account. In fact, the
presence of the sharp re-entering edge at the connection of the
riser to the casting increases the cooling rate, and consequently
leads to a cold spot-like region. Note that in original specimens of
Niyama, this a conical connection is considered (not with a sudden
cross section), and as a results such ring-like regions were not
detected in their analysis.


\clearpage

\subsection{Industrial hammer casting. }%
\label{sec:hammer_casting}%

As a more realistic example, we consider the solidification of a
steel hammer casting used for the 1988 TMS-AIME conference as a
benchmark \citep{george1988tpm}. Detailed information about
geometries, physical properties and casting conditions can be found
in \citep{george1988tpm,tavakoli2007efd}. The hammer's mold-box is
discretized into a uniform Cartesian grid with mesh size 3 mm, using
the Spen Source software package CartGen
\citep{tavakoli2007cartgen}\footnote{\href{http://mehr.sharif.ir/~tav/cartgen.htm}{CartGen:
http://mehr.sharif.ir/$\tilde{\ }$tav/cartgen.htm}}, resulting in
about 6 millions grid points.

\begin{figure}[ht]%
\begin{center}%
\ifshowfigs%
\includegraphics[width=15.cm]{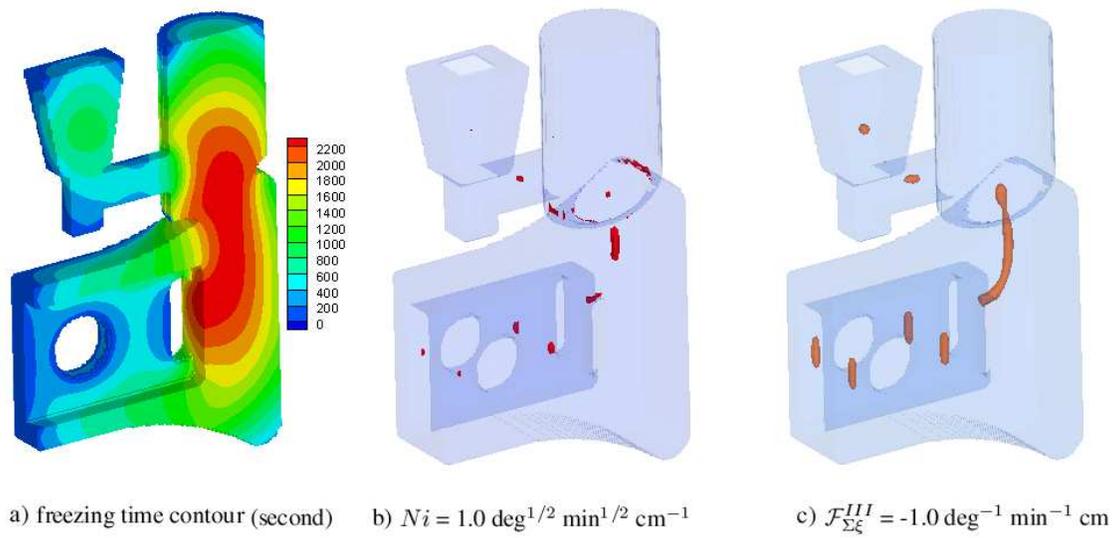}%
\fi%
\caption{Results of simulations for industrial hammer casting: %
a)~Local freezing time contours at a selected section (in seconds), %
b)~Iso-contour of the critical value $Ni$ = 1.0 deg$^{1/2}$ min$^{1/2}$
cm$^{-1}$ of the Niyama criterion,
c)~Iso-contour of
$\mathcal{F}_{\Sigma\xi}^{III}$ = -1.0 deg$^{-1}$ min$^{-1}$ cm.}%
\label{fig:hammer}%
\end{center}%
\end{figure}

\autoref{fig:hammer} shows the results of this numerical experiment,
including local freezing time contours, and iso-contours at critical
value for the Niyama criterion and $\mathcal{F}_{\Sigma\xi}^{III} =$
-1.0  deg$^{-1}$ min$^{-1}$ cm. The plot shows that both methods
predict the correct location of the major hot spot below the riser
as well as four separated hot spots around the hammer's core.
However, the defect contrast and extent of predicted macroshrinkage
provided by the suggested measure in this study is better than by
the Niyama criterion. It is obvious that the Niyama criterion hardly
detects the macroshrinkage which is formed in the pouring basin due
to the formation of an isolated liquid pool at the early stage of
the solidification there while the criterion proposed herein detects
this defect well. Finally, Niyama's criterion predicts some
fictitious defect regions due to the violation of the necessary
condition mentioned in \autoref{sec:analysisofcfm}.\\

\clearpage

\section{Closing remarks. }%
\label{sec:closing-remarks}%

The prediction of the centerline shrinkage in metal castings using
thermal criteria is a practically very important problem. In this
contribution, we have briefly reviewed existing methods and analyzed
their important limitations. Our analysis results that thermal
criteria are in fact macro-shrinkage prediction measure not unlike
available miss-leadings in literature which introduce them as
micro-shrinkage prediction measures. Moreover, application of these
measures are limited to short freezing range alloys and when the
quality of melt is good. Our analysis also suggests that that
critical values for thermal criteria functions are essentially
shape-dependent and these measures should be more considered as
qualitative measures than quantitative ones.

Based on a heuristic two-scale analysis, a new thermal criterion
function is introduced. Our main motivation was moderating the
shape-dependency problem of the critical values for thermal criteri
functions, in light provided by our analysis. The feasibility of the
suggested measure is shown using available experimental data. Our
results show that the measure $\kappa v_s G^{-1}$ with a threshold
values around -0.5 to -10.0 deg$^{-1}$ min$^{-1}$ cm is success in
predicting the formation of centerline shrinkages in steel castings.
This measure can be generalized to the the form $\big( a + b \kappa
\big) v_s^c G^d$. Obviously, experiments are required to determine
the unknown coefficients $a$, $b$, $c$ and $d$ as well as
corresponding critical thresholds for specific alloy and casting
conditions.

\begin{figure}[ht]%
\centerline{\includegraphics[width=12cm]{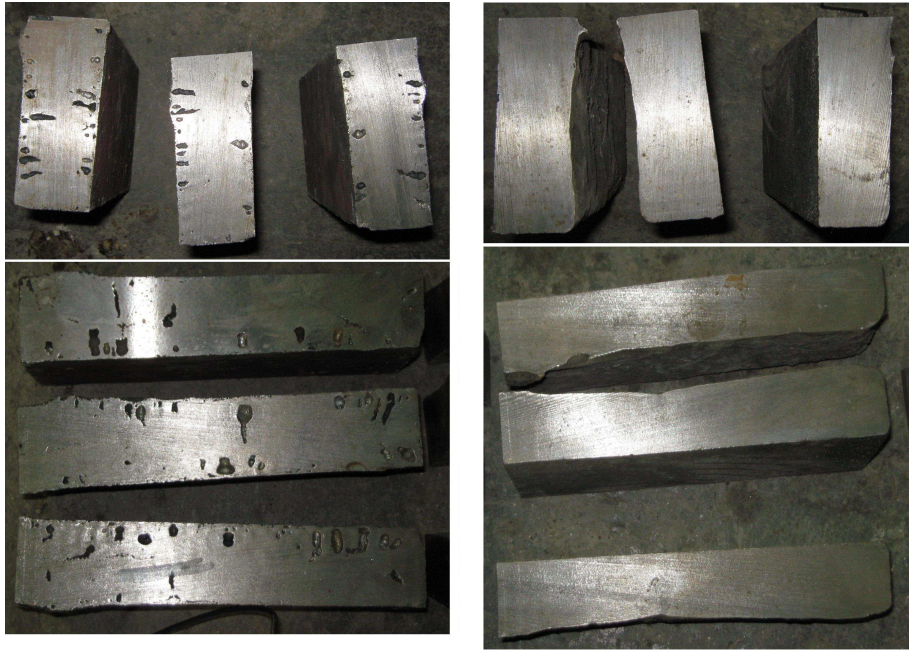}}%
\caption{Cross section of CK45 plate-like (40$\times$20$\times$4 cm)
steel castings incompletely de-gassed melt (left) and well de-gassed
melt (right).
}%
\label{fig:rt_steel_plate}%
\end{figure}

{\bf\csection \noindent Appendix: Formation of shrinkage in
plate-like castings. } In this section we provide some experimental
data which support our a part of our argument in this study. our
experiments are related to casting of 40$\times$20$\times$4 cm
plate-like CK45 carbon steel. We cast our sample without any feeding
and just design gating system such that it (hopefully) compensates
liquid-state shrinkage of melt. couple of samples are cast under two
conditions: 1) not sufficiently de-gassed melt and well de-gassed
melt (with 0.1 weight percent aluminium).
\autoref{fig:rt_steel_plate} show cross section of these castings.
As plots show, for melt with dissolved gas shrinkages are combined
with gas porosity and are distributed within cast. It should be
noted that regions near mold walls are free of defects in this case.
according to plot, there is no sign of shrinkage within cast for the
case of well de-gased melt. Instead, the shrinkage is compensated by
surface sink in this case. We partly connect this observation to the
difficulty of defect nucleation in a clean melt and possibility of
casting surface deformation for plate-like castings. A part of this
observation can be connected to effect of gravity too.

\begin{figure}[ht]%
\begin{center}%
\ifshowfigs%
\includegraphics[width=10.cm]{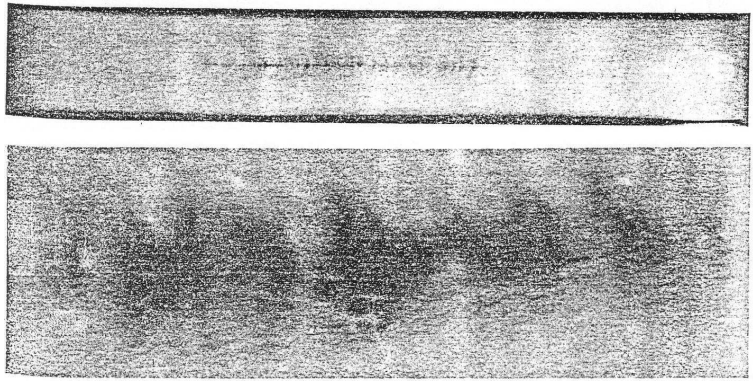}%
\fi%
\caption{X-ray photographs related to figure 11 (bottom) and figure
17 (top) of \citet{pellini1953fwd}
paper which are related to plate and bar castings respectively.}%
\label{fig:pellini_xray}%
\end{center}%
\end{figure}

By this experiment, we want to conclude that formation of central
shrinkage in plate-like castings are very difficult if the quality
of melt be high. And if shrinkages are observed in central regions
they are more like dispersed shrinkages which nucleate with the aid
of pre-existing bubble and/or supersaturation of melt with dissolved
gases. To justify this prediction, we show X-ray photograph of
Pellini for his plate- and bar- like samples in
\autoref{fig:pellini_xray}. Plot clearly show that shrinkage
appeared enterally for bar-like sample while it has appearance of
dispersed micro-porosities for the case of plate-like sample which
is a sign for low quality melt.

With the same reasoning, we want to make doubt on validity
experimental works reported in \cite{carlson2002dnf,ou2002dnf} as
they used plat-like (also bar-like which is not in question) samples
and only used X-ray photography to detect defect regions. They did
not report any photograph from cross-section of their castings which
determines type of their defects (dispersed microporosity due to
melt preparing procedure or actual concentrated center-line
shrinkages). In particular their casting are performed by couple of
industrial foundations which increase susceptibility of careless
melt preparation.

{\bf\csection \noindent Acknowledgment. } The authors would like to
thank Wolfgang Bangerth which contribute to early version of this
paper. Experimental part of our work was supported by Razi
Metallurgical Research Center, their support is greatly
acknowledged.

\single%

\bibliography{biblio}%

\end{document}